\documentclass[preprint]{ptephy_v1}

\preprintnumber{XXXX-XXXX} 
\usepackage{hyperref}
\usepackage{ulem}
\usepackage{color}

\begin{document}

\title{On the mapping after and before truncation in the boson expansion theory}


\author{Kimikazu Taniguchi}
\affil{Department of Health Data Science, Suzuka University of Medical Science, Suzuka  1001-1, Japan\email{kimikazu@suzuka-u.ac.jp}}





\begin{abstract}
Using the norm operator method, which extends and corrects the conventional boson expansion theories, we investigate two boson mappings of the boson expansion theory, the so-called mapping after truncation and the mapping before truncation.
The difference between them stems from the treatment of the phonon excitation modes; those not adopted as boson excitation modes in the former mapping are {first all} adopted {as boson excitation modes and then truncated later} in the latter mapping.
{If and only if }the commutation relations among the phonon operators are closed among the excitation modes adopted as the boson excitation modes in the mapping after truncation, the mapping after and the mapping before truncation coincide, not depending on the types, Hermitian and non-Hermitian.
We also investigate the Park operator, which judges whether a boson state vector is physical, and reveal that the conventional claim, which claims that it is applicable only when the mapping is that of the whole fermion space, is incorrect.
\end{abstract}

\subjectindex{xxxx, xxx}

\maketitle

\section{Introduction}
The boson expansion theory is one of the many-body theories to elucidate the large amplitude motions of atomic nuclei \cite{KM91}.

The boson expansion theory can be formulated by the one-to-one mappings of the fermion space that consists of the even number of quasi-particles or its subspace into the boson subspace \cite{Usi60, MYT64, LH75, KT83, Ta01}.
These mappings embed the effect of the Pauli exclusion principle into the boson subspace.
The boson operators obtained by mapping are expressed by the expansion form, which embodies the effect of the Pauli exclusion principle, and the boson state vectors that correspond to the fermion ones also generally reflect the effect of the Pauli exclusion principle.

The mapping for the whole fermion space (MWFS) corresponds all of each pair excitation of the quasi-particle to each excitation of the boson.
The basis of the whole fermion space with the even number of the quasi-particles and the completely anti-symmetrized boson state vector correspond to one another \cite{Usi60, MYT64}.
Any state vector of the fermion space is mapped to 
the superposition of these completely anti-symmetrized boson state vectors.

On the other hand, the mapping after truncation(MAT) \cite{LH75} maps only the fermion subspace constructed by the selected phonons, to make the effect of the Pauli principle as weak as possible, onto the boson subspace whose bosons have the same excitation modes as the selected phonons.
As a result, it becomes possible, by the proper selection of the excitation modes and the limitation of the number of the excitation, that the ideal boson state vectors, which reflect no effect of the Pauli exclusion principle, become physical state vectors, which correspond to the fermion state vectors, and the effect of the Pauli principle reflects itself only to the mapped operator{s} \cite{KT83}. 
The conventional practical boson expansion theories adopt MAT \cite{KT83, Ta01}.

Although it is claimed that the mapping operator of MAT cannot be obtained by {the mapping before truncation (MBT) \cite{LH75}, which merely truncates} the boson excitation modes of MWFS that correspond with the non-adopted phonon excitation modes in MAT, the same expansions are obtained by MBT and MAT in the Dyson boson expansion theory (DBET){.} Since the boson expansions are finite in DBET, the operators of MAT and MBT are identical,
{which suggests that in DBET, MBT coincides with MAT.

There is the Park operator, which determines whether a state vector is physical for MWFS \cite{P87}. Since this operator is DBET's, that of MAT is obtained by MBT.  If MBT coincides with MAT, the Park operator should also be able to properly judge whether the state vectors used as MAT are physical.}
The following is, however, claimed:

{\it This Park's idea seems quite elaborate, but one should notice that the Park operator is meaningful if and only if it is applied to the whole boson space. If the Park operator is used in some truncated boson subspace, then its eigenvalue drastically changes depending on the degree of truncation. Thus, the Park operator works only for the case of boson mapping for the whole fermion space} \cite{Ta01}.

DBET derives the boson expansions under the assumption that the closed-algebra approximation holds, which holds by the non-adopted modes discarding (NAMD){\cite{NR23}}.
KT-3, which is of Hermitian type with infinite expansions, also adopts NADM \cite{KT83}.
The norm operator method, which is a new version of the boson expansion theories, however, has revealed that the conventional claims on NADM \cite{KT83, SK88, Ta01} are incorrect \cite{NR23}.
{The conventional improper interpretation of NAMD could have made the above claim for the Park operator.}
Therefore, {the norm operator method should be used to find the relation between MAT and MBT and whether the conventional claim about the Park operator is correct.}

{This paper re-examines} MBT and the Park operator using the norm operator method.

In section \ref{fsbs}, the notations and the definitions of the fermions and the bosons are given.

In section \ref{norm}, we offer an outline of the formulation of the norm operator method and some relevant results obtained.

In section \ref{mbt}, we investigate the mapping before truncation and the Park operator by the norm operator method.

Section \ref{sum} is a summary.

\section{Fermion space and boson space}
\label{fsbs}
In this section, the notations and the definitions of the fermions and the bosons are given.

We introduce the Tamm-Dancoff phonon creation and anihilation operators,
\begin{subequations}
\label{eq:phononop}
\begin{equation}
\label{eq:phononopc}
X_\mu^\dagger
=\displaystyle\sum_{\alpha<\beta}\psi_\mu(\alpha\beta)a_\alpha^\dagger
a_\beta^\dagger,
\end{equation}
\begin{equation}
\label{eq:phononopa}
X_\mu=\displaystyle\sum_{\alpha<\beta}\psi_\mu(\alpha\beta)a_\beta
a_\alpha.
\end{equation}
\end{subequations}
Here, $a_\alpha^\dagger$ and $a_\alpha$ are quasi-particle creation and annihilation operators in a single-particle state $\alpha$.
The coefficients satisfy the following relations:
\begin{subequations}
\label{eq:TDrel}
\begin{equation}
\label{eq:anti}
\psi_\mu(\beta\alpha)=-\psi_\mu(\alpha\beta)
\end{equation}
\begin{equation}
\label{eq:orthonormal}
\sum_{\alpha<\beta}\psi_\mu(\alpha\beta)\psi_{\mu'}(\alpha\beta)=\delta_{\mu,
\mu'},
\end{equation}
\begin{equation}
\label{eq:complete}
\sum_{\mu}\psi_\mu(\alpha\beta)\psi_{\mu}(\alpha'\beta')=\delta_{\alpha,
\alpha'}\delta_{\beta, \beta'}-\delta_{\alpha,
\beta'}\delta_{\beta, \alpha'},
\end{equation}
\end{subequations}

The phonon operators satisfy the following commutation relations:
\begin{subequations}
\label{eq:algebra}
\begin{equation}
\label{eq:algebra1}
[ X_\mu, X_{\mu'}^\dagger ]=\delta_{\mu,
\mu'}-\sum_q\Gamma^{\mu\mu'}_qB_q,
\end{equation}

\begin{equation}
\label{eq:algebra2}
[ B_q, X_\mu^\dagger
]=\sum_{\mu'}\Gamma^{\mu\mu'}_qX_{\mu'}^\dagger,
\end{equation}

\begin{equation}
\label{eq:algebra3}
[ X_\mu, B_q ]=\sum_{\mu'}\Gamma^{\mu'\mu}_qX_{\mu'},
\end{equation}
\end{subequations}
where the definition of $\Gamma^{\mu\mu'}_q$ is as follows:
\begin{equation}
\label{eq:Gamma}
\Gamma^{\mu\mu'}_q=\sum_{\alpha\beta}\varphi_q(\alpha\beta)\Gamma^{\mu\mu'}_{\alpha\beta},\quad
\Gamma^{\mu\mu'}_{\alpha\beta}=\sum_\gamma\psi_\mu(\alpha\gamma)\psi_{\mu'}(\beta\gamma).
\end{equation}
The following relation holds:
\begin{equation}
\label{eq:qbargam}
\Gamma_{\bar q}^{\mu_1 \mu_2}=\Gamma_q^{\mu_2 \mu_1}.
\end{equation}
$B_q$, the so-called scattering operators, are defined as
\begin{subequations}
\label{eq:scop12}
\begin{equation}
\label{eq:scop}
B_q=\sum_{\alpha\beta}\varphi_q(\alpha\beta)a_\beta^\dagger
a_\alpha,
\end{equation}
\begin{equation}
\label{eq:scop2}
B_{\bar q=}B_q^\dagger,
\end{equation}
\end{subequations}
where
\begin{subequations}
\label{eq:coeffscop}
\begin{equation}
\label{eq:barq}
\varphi_{\bar q}(\alpha\beta)=\varphi_q(\beta\alpha),
\end{equation}
\begin{equation}
\label{eq:coeffscop1}
\displaystyle\sum_{\alpha\beta}\varphi_q(\alpha\beta)\varphi_{q'}(\alpha\beta)=\delta_{q,q'},
\end{equation}
\begin{equation}
\label{eq:coeffscop2}
\displaystyle\sum_{q}\varphi_q(\alpha\beta)\varphi_{q}(\alpha'\beta')=\delta_{\alpha,\alpha'}\delta_{\beta,
\beta'}.
\end{equation}
\end{subequations}

From Eqs. (\ref{eq:algebra1}) and (\ref{eq:algebra2}), we obtain
\begin{equation}
\label{eq:doublecom}
[ [X_{\mu_1}, X_{\mu_2}^\dagger], X_{\mu_3}^\dagger] = -\sum_{\mu'}Y(\mu_1 \mu_2 \mu_3 \mu')X_{\mu'}^\dagger,
\end{equation}
where the definition of $Y(\mu_1\mu_2\mu_3\mu_4)$ is 
\begin{equation}
\label{eq:Y}
Y(\mu_1\mu_2\mu_3\mu_4)=\sum_q\Gamma_q^{\mu_1\mu_2}\Gamma_q^{\mu_3\mu_4}
=\sum_{\alpha\beta}\Gamma_{\alpha\beta}^{\mu_1\mu_2}\Gamma_{\alpha\beta}^{\mu_3\mu_4}.
\end{equation}
The following relation holds:
\begin{equation}
\label{eq:Ysym}
\begin{array}{lll}
Y(\mu_1\mu'_1\mu'_2\mu_2)
&=&Y(\mu_2\mu'_1\mu'_2\mu_1)
\\
&=&Y(\mu_1\mu'_2\mu'_1\mu_2)
\\
&=&Y(\mu'_1\mu_1\mu_2\mu'_2).
\\
\end{array}
\end{equation}

Correspondingly to the phonons, bosons are introduced,
\begin{equation}
\label{eq:bcom}
[ b_\mu, b_{\mu'}^\dagger ] = \delta_{\mu, \mu'}.
\end{equation}
The multi-boson state vectors,
\begin{equation}
\label{eq:multib}
\vert N; \mu))=\vert \mu_1, \mu_2, \cdots , \mu_N))=b_{\mu_1}^\dagger
b_{\mu_2}^\dagger \cdots b_{\mu_N}^\dagger \vert 0),
\end{equation}
are orthogonal to one another, and are normalized by their norms,
\begin{equation}
\label{eq:normb}
\mathcal{N}_B(N; \mu)=((N: t\vert N; \mu)),
\end{equation}
such as
\begin{equation}
\label{eq:bbasis}
\vert N; \mu)=\vert \mu_1, \mu_2, \cdots , \mu_N)=\mathcal{N}_B(N; \mu)^{-1/2}\vert N; \mu)).
\end{equation}
They are so-called ideal boson state vectors.
$\vert N:\mu)$ can be an orthnormal basis of the boson space.

While the multi-phonon states,
\begin{subequations}
\label{eq:mulphs}
\begin{equation}
\label{eq:mulphs1}
\vert N; \mu\rangle\rangle=\vert \mu_1, \mu_2, \cdots ,
\mu_N\rangle\rangle=X_{\mu_1}^\dagger X_{\mu_2}^\dagger \cdots
X_{\mu_N}^\dagger \vert 0\rangle,
\end{equation}
\begin{equation}
\label{eq:mulphs2}
\vert N;\mu\rangle=\mathcal{N}_B(N; \mu)^{-\frac12}\vert N;\mu\rangle\rangle,
\end{equation}
\end{subequations}
are not orthonormal and are even overcomplete at the stage of $N=2$ with all phonon modes adopted \cite{KT83}. 
These multi-phonon state vectors span a fermion space consisting of an even number of quasi-particles.
Its orthonormal bases are nothing but the following:
\begin{equation}
\label{fbasis}
\vert\alpha_1\beta_1\alpha_2\beta_2\cdots\alpha_N\beta_N\rangle=a_{\alpha_1}^\dagger a_{\beta_1}^\dagger a_{\alpha_2}^\dagger a_{\beta_2}^\dagger\cdots a_{\alpha_N}^\dagger a_{\beta_N}^\dagger\vert 0\rangle.
\end{equation}
The multi-phonon state vectors,
\begin{equation}
\label{eq:mulphst}
\vert N; t\rangle\rangle=\vert t_1, t_2, \cdots ,
t_N\rangle\rangle=X_{t_1}^\dagger X_{t_2}^\dagger \cdots
X_{t_N}^\dagger \vert 0\rangle\quad (0\leq N\leq N_{max}),
\end{equation}
with restricting the phonon excitation modes and the number of phonon excitations, span its subspace.
$\{t\}$ usually consists of collective modes and some non-collective modes if necessary, selected by the small amplitude approximation. 
Hereafter, we denote the excitation modes other than $\{t\}$ as $\{\bar t\}$.

\section{The norm operator method}
\label{norm}
In this section,  we offer an outline of the formulation of the norm operator method and some relevant results obtained.

The mapping operator of the norm operator method is formulated as MAT, which restricts not only the types of phonon excitation modes but also the number of excitations. By removing these restrictions, MWFS can also be treated.

The mapping operator is
\begin{equation}
\label{eq:bmopzubar}
U_\xi=\hat Z^{\xi-\frac 12}\widetilde U,
\end{equation}
where $\widetilde U$,
\begin{subequations}
\label{eq:tildempop}
\begin{equation}
\label{eq:tildempop1}
\widetilde {U}=\sum_{N=0}^{N_{max}}{\widetilde U}(N),
\end{equation}
\begin{equation}
\label{eq:tildempop2}
\begin{array}{lll}
{\widetilde U}(N)&=&\displaystyle\sum_{t}\vert N; t)\langle N; t\vert
\\
&=&
\displaystyle\sum_{t_1\leq t_2\leq\cdots\leq t_N}\vert t_1 t_2\cdots
t_N)\langle t_1 t_2\cdots t_N\vert,
\end{array}
\end{equation}
\end{subequations}
and $\hat Z$,
\begin{subequations}
\begin{equation}
\label{eq:normop}
\hat Z=\sum_{N=0}^{N_{max}}\hat Z(N),
\end{equation}
\begin{equation}
\label{eq:normop2}
\begin{array}{lll}
\hat{Z}(N)&=&\displaystyle\sum_{t t'}\vert N, t)\langle N;
t\vert N; t'\rangle ( N; t' \vert
\\
&=&
\displaystyle\sum_{t_1\leq\cdots \leq t_N}\sum_{t'_1\leq\cdots
\leq t'_N}\vert t_1 \cdots t_N)\langle t_1\cdots t_N\vert
t'_1\cdots t_N\rangle (t'_1\cdots t'_N \vert,
\end{array}
\end{equation}
\end{subequations}
is a norm operator.
We denote the eigenvalues and eigenvectors of $\hat Z(N)$ as $\hat z_a(N)$ and $\vert N; a)$, respectively.
Using these, the mapping operator is expressed as
\begin{equation}
\label{eq:bmop}
U_\xi=\sum_{N=0}^{N_{max}}U_\xi(N);\quad U_\xi(N)=\sum_{a\neq a_0}z_a(N)^\xi\vert N; a)\langle N; a\vert,
\end{equation}
where $\vert N; a_0)$ denotes an eigenvector whose eigenvalue is zero, and $\vert N; a\rangle (a\neq a_0)$,
\begin{equation}
\label{eq:orthonormbs}
\vert N; a\rangle = {z_a(N)}^{-\frac 12}\sum_t(N;t\vert N;a)\vert N; t\rangle
\qquad (a\neq a_0),
\end{equation}
are normalized orthogonal vectors constructed from the multi-phonon state vectors.
The following relations are satisfied:
\begin{equation}
\label{eq:utftbh}
U_{-\xi}^\dagger U_{\xi}=\hat T_F,\qquad U_{\xi}U_{-\xi}^\dagger =\hat T_B,
\end{equation}
where
\begin{subequations}
\begin{equation}
\label{eq:unitf}
\hat T_F=\sum_{N=0}^{N_{max}}\hat T_F(N);\qquad
\hat T_F(N)=\displaystyle\sum_{a\neq a_0}\vert N; a\rangle\langle N; a\vert,
\end{equation}
\begin{equation}
\hat T_B=\sum_{N=0}^{N_{max}}\hat T_B(N);\quad\hat T_B(N)=\sum_{a\neq a_0}\vert N; a)(N; a\vert.
\end{equation}
\end{subequations}
We introduce the following operators,
\begin{equation}
\label{eq:breve1B}
\breve 1_B=\sum_{N=0}^{N_{max}}\hat 1_B(N);\qquad \hat 1_B(N)=\sum_t\vert N; t)(N;t\vert.
\end{equation}
If $\hat Z(N)$ has even one zero eigenvalue, then $\hat T_B(N)\neq \hat 1_B(N)$ and hence $\hat T_B\neq \breve 1_B$. Otherwise, $\hat T_B(N)= \hat 1_B(N)$ and $\hat T_B=\breve 1_B$, which is realized under the proper limitation of the sort of the phonon excitation modes and the number of the phonon excitations.
The norm operator $\hat Z$ satisfies the relation,
\begin{equation}
\label{eq:zzinv}
\hat Z^\xi\hat Z^{-\xi}=\hat Z^{-\xi}\hat Z^\xi=\hat T_B.
\end{equation}

The state vectors and operators of the fermion space are mapped onto those of the boson subspace as
\begin{subequations}
\label{eq:ximap}
\begin{equation}
\label{eq:ximap1}
\vert \psi')_{\xi}= U_{\xi}\vert\psi'\rangle,\qquad {}_{-\xi} (\psi\vert =\langle\psi\vert U_{-\xi}^\dagger,
\end{equation}
\begin{equation}
\label{eq:ximap2}
(O_F)_{\xi}=U_{\xi}O_FU_{-\xi}^\dagger.
\end{equation}
\end{subequations}
The mapping is of the Hermitian type when $\xi=0$ and, in other cases, of the non-Hermitian type.

$U_\xi$ and $U_{-\xi}^\dagger$ realize one-to-one correspondence between the fermion subspace projected by $\hat T_F$ and the boson subspace projected by $\hat T_B$.
For the state vectors, $\vert\psi\rangle$ and $\vert\psi'\rangle$, which belong to the fermion subspace projected by $\hat T_F$,
\begin{equation}
\label{eq:mteqh}
\begin{array}{lll}
\langle\psi\vert O_F\vert\psi'\rangle
&=&\langle\psi\vert\hat T_F O_F\hat T_F\vert\psi'\rangle
\\
&=&\langle\psi\vert U_{-\xi}^\dagger U_\xi O_FU_{-\xi}^\dagger U\vert\psi'\rangle
\\
&=&{}_{-\xi}(\psi\vert(O_F)_\xi\vert\psi')_\xi,
\end{array}
\end{equation}
that is, the matrix element of the fermion subspace becomes equal to that of the corresponding boson subspace. 
The boson subspace corresponding to the fermion subspace projected by $\hat T_F$ is called the physical subspace, and the boson state vectors belonging to that space are called the physical state vectors. $\hat T_B$ is the projection operator onto the physical subspace.

We denote the mapping of $\widetilde U$ as
\begin{subequations}
\label{eq:tildemap}
\begin{equation}
\label{eq:tildemap1}
\widetilde{\vert \psi)}= \widetilde U\vert\psi\rangle,\qquad\widetilde{(\psi\vert} =\langle\psi\vert {\widetilde U}^\dagger,
\end{equation}
\begin{equation}
\label{eq:tildemap2}
\widetilde{O_F}=\widetilde UO_F{\widetilde U}^\dagger.
\end{equation}
\end{subequations}
The mapping of Eqs. (\ref{eq:ximap}) is expressed as
\begin{subequations}
\label{eq:ximapzbar}
\begin{equation}
\label{eq:ximapzbar1}
\vert \psi')_{\xi}= \hat Z^{\xi-\frac 12}\widetilde{\vert\psi')},\qquad {}_{-\xi} (\psi\vert=\widetilde{(\psi\vert}\hat Z^{-\xi-\frac 12},
\end{equation}
\begin{equation}
\label{eq:ximapzbar2}
(O_F)_{\xi}=\hat Z^{\xi-\frac 12}\widetilde{O_F}\hat Z^{-\xi-\frac 12},
\end{equation}
\end{subequations}
which makes it clear that the different treatment of the norm operator in the mapping operator produces another type of mapping.
{The mapping of Eqs. (\ref{eq:ximap}) is also expressed as
\begin{subequations}
\label{eq:nhhr}
\begin{equation}
\label{eq:nhhr1}
\vert \psi')_{\xi}= \hat Z^{\xi}\vert\psi')_0,\quad {}_{-\xi} (\psi\vert={}_0(\psi\vert\hat Z^{-\xi},
\end{equation}
\begin{equation}
\label{eq:nhhr2}
(O_F)_{\xi}=\hat Z^{\xi}(O_F)_0\hat Z^{-\xi},
\end{equation}
\end{subequations}
The mapping of $\xi=0$ being of the Hermitian type and that of $\xi\neq 0$ being of the non-Hermitian type transform one another by the similarity transformation operator that becomes the power of the norm operator $\hat Z$.
$\xi=\frac 12$ for obtaining DBET.}

Replacing $\{t\}$ by $\{\mu\}$, the mapping becomes MWFS with the phonon excitation number restriction.
Hereafter, we attach $(A)$ such as $\hat Z^{(A)}$ in the case that we introduce boson operators corresponding to all phonon excitation modes for no confusion.

Introducing another boson operators,
\begin{equation}
\label{eq:babbmu}
b_{\alpha\beta}=\sum_\mu\psi_\mu(\alpha\beta)b_\mu, \quad b_{\alpha\beta}^\dagger=\sum_\mu\psi_\mu(\alpha\beta)b_\mu^\dagger,
\end{equation}
which satisfies the commutation relations,
\begin{subequations}
\label{eq:babcom}
\begin{equation}
[b_{\alpha'\beta'}, b_{\alpha\beta}^\dagger]=\delta_{\alpha'\alpha}\delta_{\beta'\beta}-\delta_{\alpha'\beta}\delta_{\beta'\alpha}.
\end{equation}
\begin{equation}
[b_{\alpha'\beta'}, b_{\alpha\beta}]=0, \quad [b_{\alpha'\beta'}^\dagger, b_{\alpha\beta}^\dagger]=0.
\end{equation}
\end{subequations}
{
Using these operators, $\widetilde{U}^{(A)}(N)=\displaystyle\sum_\mu\vert N; \mu)\langle N; \mu\vert$ are rewritten as
\begin{equation}
\widetilde{U}^{(A)}(N)=\sqrt{(2N-1)!!}\sum_{\alpha_1<\beta_1<\cdots<\alpha_N<\beta_N}\vert\alpha_1\beta_1\cdots\alpha_N\beta_N)_M\ \langle\alpha_1\beta_1\cdots\alpha_N\beta_N\vert,
\end{equation}
where
\begin{equation}
\vert\alpha_1\beta_1\cdots\alpha_N\beta_N)_M=\frac{1}{\sqrt{(2N-1)!!}}{\sum_{P}}'(-)^Pb_{\alpha_1\beta_1}^\dagger\cdots b_{\alpha_N\beta_N}^\dagger\vert 0),
\end{equation}
and $\displaystyle{\sum_{P}}'$ means the summation so that the states on the left side become totally antisymmetric \cite{MYT64}.
From these, we obtain
\begin{equation}
\begin{array}{lll}
\hat Z^{(A)}(N)&=&\widetilde{U}^{(A)}(N)\widetilde{U}^{(A)}(N)^\dagger
\\
&=&\displaystyle (2N-1)!!\displaystyle\sum_{\alpha_1<\beta_1<\cdots\alpha_N\beta_N}\vert\alpha_1\beta_1\cdots\alpha_N\beta_N)_M{}_M(\alpha_1\beta_1\cdots\alpha_N\beta_N\vert,
\end{array}
\end{equation}
which is the spectral decomposition of $\hat Z^{(A)}(N)$ and indicates that the eigenvectors of the eigenvalue $(2N-1)!!$ are $\vert\alpha_1\beta_1\cdots\alpha_N\beta_N)_M$.
We also obtain
\begin{equation}
\begin{array}{lll}
\hat T_F^{(A)}&=&\displaystyle\sum_{N=0}^{N_{max}}\hat T_F^{(A)}(N),
\\
\hat T_F^{(A)}(N)&=&\hat 1_F(N),
\\
\quad\hat 1_F(N)&=&\displaystyle\sum_{\alpha_1<\beta_1<\cdots\alpha_N\beta_N}\vert\alpha_1\beta_1\cdots\alpha_N\beta_N\rangle\langle\alpha_1\beta_1\cdots\alpha_N\beta_N\vert,
\end{array}
\end{equation}
\begin{equation}
\begin{array}{lll}
\hat T_B^{(A)}&=&\displaystyle\sum_{N=0}^{N_{max}}\hat T_B(N),
\\
\hat T_B^{(A)}(N)&=&\displaystyle\sum_{\alpha_1<\beta_1<\cdots\alpha_N\beta_N}\vert\alpha_1\beta_1\cdots\alpha_N\beta_N)_M{}_M(\alpha_1\beta_1\cdots\alpha_N\beta_N\vert.
\end{array}
\end{equation}
$\hat Z^{(A)}$ is written as
\begin{equation}
\hat Z^{(A)}=(2\hat N_B^{(A)}-1)!!\hat T_B^{(A)}.
\end{equation}
}
The mapping operator is expressed as
\begin{equation}
\begin{array}{lll}
U_\xi^{(A)}&=&\displaystyle\sum_{N=0}^{N_{max}}U_\xi(N),
\\
U_\xi^{(A)}(N)&=&\displaystyle\{(2N-1)!!\}^\xi\sum_{\alpha_1<\beta_1<\cdots<\alpha_N<\beta_N}\vert\alpha_1\beta_1\cdots\alpha_N\beta_N)_M\ \langle\alpha_1\beta_1\cdots\alpha_N\beta_N\vert.
\end{array}
\end{equation}
If we set $\xi=0$ and $N_{max}\rightarrow\infty$, this mapping becomes the MYT mapping \cite{MYT64},
from which we obtain the boson expansions of Holstein and Primakoff, and if we take {$\xi=\displaystyle\frac 12$}, they become mapping operators for the Dyson boson expansions \cite{JDF71}.
Taking $N_{max}\rightarrow\infty$, $U_\xi^{(A)}$ becomes a mapping operator of MWFS.
We use $(W)$ to denote the case where all phonon excitation modes are taken as boson excitations without restricting the number of phonon excitations.
$U^{(W)}=\lim_{N_{max}\rightarrow\infty} U^{(A)}$ is the mapping operator of MWFS, and $\hat T_F^{(W)}=\lim_{N_{max}\rightarrow\infty}\hat T_F^{(A)}$ is nothing more than the unit operator $\hat 1_F$ in the fermion space consisting of even quasi-particles.
The identity operator in the boson space is $\hat 1_B=\lim_{N_{max}\rightarrow\infty}\breve 1_B$.

The boson expansion of $(O_F)_\xi$ can be obtained by those of $\hat Z$ and $\widetilde{O_F}$.
In the conventional boson expansion theories, the boson expansions are obtained by discarding, without proper consideration, the phonon excitation modes not adopted as boson excitations \cite{KT83, SK88, Ta01}.
{The norm operator method has found for the first time that correctly handling the phonon excitation modes $\{\bar t\}$ that are not adopted as boson excitations in MAT, it is not sufficient to focus only on the norm operator $\hat Z$ consisting only of the excitation modes of $\{t\}$; it is also necessary to pay attention to its relationship with the norm operator $\hat Z^{(A)}$ consisting of all phonon excitation modes $\{\mu\}$.
This enables us to establish a method to obtain a small parameter expansion with the boson approximation as the zeroth approximation without assuming NAMD.
It has also been revealed that NAMD makes the Hermitian type and all the other boson expansions essentially finite and the small parameter expansions impossible.
Although a comment \cite{S24} was made expressing objections to the norm operator method, which leads to the results that overturn the conventional wisdom of the boson expansion theories, this has been immediately and correctly refuted \cite{T24}.}

The relation between $\hat Z^{(A)}$ and $\hat Z$ is
\begin{subequations}
\begin{equation}
\hat Z^{(A)}=\hat Z+\hat W+\hat W^\dagger+\hat Z,'
\end{equation}
where
\begin{equation}
\hat Z=\breve 1_B\hat Z^{(A)}\breve 1_B,
\quad\hat W=\breve 1_B\hat Z^{(A)}(\breve 1_B^{(A)}-\breve 1_B),
\quad\hat Z'=(\breve 1_B^{(A)}-\breve 1_B)\hat Z^{(A)}(\breve 1_B^{(A)}-\breve 1_B).
\end{equation}
\end{subequations}
In the case that NAMD holds, the phonon commutation relations are closed:
\begin{equation}
\label{eq:clarg}
[ [X_{t_3}, X_{t_1}^\dagger], X_{t_2}^\dagger] = -\sum_{t'}Y(t_3t _1t_2 t')X_{t'}^\dagger,
\end{equation}
These holding are necessary and sufficient conditions for $Y(t_3t_1t_2\bar t')=0$, i.e., $\langle t_1t_2\vert t_3 \bar t'\rangle =0$.
And, If $\langle t_1t_2\vert t\bar t\rangle =0$ hold for any $t_1$, $t_2$, $t$, and $\bar t$, $\langle t_1t_2\vert \bar t_1\bar t_2\rangle =0$ hold for any $t_1$, $t_2$, $\bar t_1$, and $\bar t_2$, and then $\hat W(2)=0$, and if $\hat W(2)=0$, $\hat W=0$. It is nothing but NAMD holds.
Therefore, the commutation relation of the phonon operators being closed within $\{t\}$, $\hat W=0$, and NAMD holding are necessary and sufficient one another.
On the other hand, regardless of what $\hat Z$, $\hat W$, and $\hat Z'$ are,
\begin{equation}
\label{eq:normopA}
\hat Z^{(A)}=(2\hat N_B^{(A)}-1)!!\hat T_B^{(A)};\quad\hat N_B^{(A)}=\sum_\mu b_\mu^\dagger b_\mu,
\end{equation}
holds.
And if $\hat W=0$ holds, then
\begin{subequations}
\begin{equation}
\hat Z=\breve 1_B\hat Z^{(A)}=\hat Z^{(A)}\breve 1_B=\breve 1_B\hat Z^{(A)}\breve 1_B,
\end{equation}
holds, and we obtain
\begin{equation}
\hat T_B=\breve 1_B\hat T_B^{(A)}=\hat T_B^{(A)}\breve 1_B=\breve 1_B \hat T_B^{(A)}\breve 1_B.
\end{equation}
\end{subequations}
Finally,
\begin{equation}
\label{eq:znnons2}
\hat Z=(2\hat N_B-1)!!\hat T_B;\quad\hat N_B =\sum_t b_t^\dagger b_t
\end{equation}
is derived.
This shows that the small parameter expansion where $\breve 1_B$ is the zeroth approximation for $\hat Z$ is impossible.
Furthermore, when $O_F$ is the phonon creation operator, anhilation operator, or scattering operator and its Dyson boson expansion is $(O_F)_D$,
\begin{equation}
\label{eq:oz}
\widetilde{O_F}=(O_F)_D\hat Z,
\end{equation}
hold, where
\begin{subequations}
\label{eq:dbexp}
\begin{equation}
(X_{t'})_D=b_{t'},
\end{equation}
\begin{equation}
\label{eq:xdaggerd}
(X_t^\dagger)_D=b_t^\dagger-\frac 12\sum_{t_1t_2}\sum_{t'_1}Y(tt_1t_2t'_1)b_{t_1}^\dagger b_{t_2}^\dagger b_{t'_1},
\end{equation}
\begin{equation}
\label{eq:bqd}
(B_q)_D=\sum_t\sum_{t'}\Gamma_q^{t't}b_t^\dagger b_{t'}.
\end{equation}
\end{subequations}
From Eq. (\ref{eq:ximapzbar2}), the boson mapping $(O_F)_\xi$ is expressed as
\begin{equation}
\label{eq:ofxi1}
(O_F)_\xi=\hat Z^{\xi-\frac 12}(O_F)_D\hat Z^{-\xi+\frac 12},
\end{equation}
which indicates that as long as it acts on a physical state vector that is an eigenstate of the boson excitation number operator, it effectively becomes a finite expansion.
For $O_F$ and $O'_F$ being the phonon operators or the scattering operators, respectively, 
\begin{subequations}
\label{eq:ofofd}
\begin{equation}
\label{eq:oftfofd}
O_FO'_F=O_F\hat T_FO'_F,
\end{equation}
holds by NAMD.
Therefore
\begin{equation}
\label{eq:ofxi2ofof}
(O_FO'_F)_\xi=(O_F)_\xi(O'_F)_\xi,
\end{equation}
and
\begin{equation}
\label{eq:ofxi2}
(O_FO'_F)_\xi=\hat Z^{\xi-\frac 12}(O_F)_D\hat T_B(O'_F)_D\hat Z^{-\xi+\frac 12},
\end{equation}
\end{subequations}
are satisfied.
In the case that $\hat W\neq 0$, Eq. (\ref{eq:oz}) and Eq. (\ref{eq:oftfofd}) do not hold.

It is clear from the above that the equations that NAMD makes hold in MAT also hold when replaced by MWFS, in which the phonon exchange relationship is closed, and in the case with only the restriction of the number of the phonon excitations.
Conversely, by applying $\breve 1_B$ to the equation that holds in MWFS, we can derive the equation that holds in {MAT} when NAMD holds.

\section{The investigation of  the mapping before truncation and the Park operator by the norm operator method}
\label{mbt}
In this section, using the norm operator method, we investigate MBT, its relation to MAT, and the Park operator.

In investigating MBT and MAT, we consider {$U_\xi^{(A)}$}, which restricts only the number of phonon excitations, and $U_\xi$, which restricts the phonon excitation modes in addition.
The MWFS is straightforwardly obtained by removing the restriction on the number of the phonon excitations of  {$U_\xi^{(A)}$}.
MBT and MAT are realized by the mapping operator $\breve 1_B^{(A)}U_\xi^{(A)}$ and $U_\xi$ respectively.

MBT restricts the physical subspace obtained by MWFS to the subspace with the specific boson excitation modes. On the other hand, MAT introduces the phonons and maps the fermion subspace spanned by the multi-phonon state vectors with the restricted excitation modes onto the physical subspace consisting only of the bosons corresponding to the phonons that construct the multi-phonon state vectors.

DBET uses the closed-algebra approximation to obtain the expansions.
It is necessary and sufficient that the closed-algebra approximation, $\hat W=0$ and NAMD hold respectively.
If $\hat W=0$, then $\breve 1_B\hat Z^{(A)}=\hat Z$, and since $[\hat Z^{(A)}, \breve 1_B]=0$,
\begin{equation}
 \breve 1_BU_\xi^{(A)}=\breve 1_B(\hat Z^{(A)})^{\xi-\frac 12}\widetilde U^{(A)}=(\breve 1_B\hat Z^{(A)})^{\xi-\frac 12}\widetilde U^{(A)}=\hat Z^{\xi-\frac 12}\widetilde U^{(A)}=\hat Z^{\xi-\frac 12}\widetilde U=U_\xi
\end{equation}
holds.
As a result, we obtain
\begin{subequations}
\label{eq:uabmapping}
\begin{equation}
\breve 1_B(O_F)_\xi^{(A)}\breve 1_B=(O_F)_\xi, 
\end{equation}
\begin{equation}
\breve 1_B\vert \psi')_\xi^{(A)}=\vert \psi')_\xi, \quad {}^{(A)}{}_{-\xi}(\psi\vert \breve 1_B={}_{-\xi}(\psi\vert,
\end{equation}
\end{subequations}
which indicates that MBT and MAT are equivalent.
This result does not depend on any particular $\xi$, and therefore also holds for the Hermitian expansions, which become finite by NAMD.

This agreement, however, cannot be obtained with small parameter expansions, which do not hold under NAMD.
When NAMD does not hold, the commutation relations of the phonon creation and annihilation operators do not close in $\{t\}$.
In {MWFS and therefore }MBT, the contribution of the phonon excitation mode $\{\bar t\}$ is borne by the boson excitations, while in MAT, it is included in the coefficients of the boson expansions.
Regardless of whether NAMD holds, MAT maps the fermion subspace projected by $\hat T_F$ to the boson subspace projected by $\hat T_B$. Further, by appropriately restricting the type and number of phonon excitation modes, $\hat T_B=\breve 1_B$ and the ideal boson state vector $\vert N; t)$ becomes physical.
On the other hand, in $U_\xi^{(A)}$, the contribution of $\{\bar t\}$ is expressed as boson excitations, so the mapping of $\hat T_F$ does not coincide with $\hat T_B$:
\begin{equation}
\label{eq:uatfua}
U_\xi^{(A)}\hat T_F{U_{-\xi}^{(A)}}^\dagger \neq \hat T_B.
\end{equation}
{The operator of the lefthandside of Eq. (\ref{eq:uatfua}) is a projection operator because
\begin{equation}
(U_\xi^{(A)}\hat T_F{U_{-\xi}^{(A)}}^\dagger )^2=U_\xi^{(A)}\hat T_F\hat T_F^{(A)}{U_{-\xi}^{(A)}}^\dagger =U_\xi^{(A)}\hat T_F{U_{-\xi}^{(A)}}^\dagger,
\end{equation}
and the following holds,
\begin{equation}
{U_{-\xi}^{(A)}}^\dagger (U_\xi^{(A)}\hat T_F{U_{-\xi}^{(A)}}^\dagger )U_\xi^{(A)}=\hat T_F^{(A)}\hat T_F\hat T_F^{(A)}=\hat T_F.
\end{equation}
Therefore, the boson subspace projected by $U_\xi^{(A)}\hat T_B{U_{-\xi}^{(A)}}^\dagger$ and the fermion subspace projected by $\hat T_F$ have one-to-one correspondence.
Thus, the physical subspace in MWFS and MAT that corresponds to the fermion subspace projected by $\hat T_F$ is different from each other.}
From Eq. (\ref{eq:uatfua})
\begin{equation}
\label{eq:1uatfua1}
\breve 1_BU_\xi^{(A)}\hat T_F{U_{-\xi}^{(A)}}^\dagger\breve 1_B\neq \hat T_B,
\end{equation}
{which indicates that the difference between MWFS and MAT in Eq. (\ref{eq:uatfua}) cannot be resolved by MBT.
It is because the physical subspace of MWFS is composed of a complex intertwining of the boson excitation modes adopted and non-adopted in MAT, and the physical subspace cannot be obtained by merely restricting it to the excitation modes adopted in MAT.
Furthermore, due to the operator $\breve 1_B$, which restricts the boson excitation modes, the left-hand side is no longer a projection operator. MBT loses the one-to-one correspondence with the fermion subspace projected by $\hat T_F$.}
Therefore, even if we restrict the types and number of phonon excitation modes so that $\hat T_B=\breve 1_B$, if NAMD does not hold, in MBT,
\begin{equation}
\label{eq:1uatfua1}
\breve 1_BU_\xi^{(A)}\hat T_F{U_{-\xi}^{(A)}}^\dagger\breve 1_B\neq \breve 1_B,
\end{equation}
{and} the ideal boson state vectors are not physical in MBT.

The mapping of $\hat T_F$ in MBT coincides with $\hat T_B$ if and only if NAMD holds, and the physical subspace of MAT that $\hat T_B$ projects is also that of MBT. 
{From these, it is also clear that MWFS maps $\hat T_F$ to $\hat T_B$ when NAMD holds.}
Furthermore, if $\hat T_B=\breve 1_B$, the ideal boson states $\vert N; t)$ become physical also in MBT {and MWFS}.

In MWFS, Park proposes an operator to determine whether a boson state vector is physical \cite{P87}.
{Using} the above analysis, it is also derived that the Park operator judges the state vectors in MBT properly {if and only if} NAMD holds.

The Park operator is defined as
\begin{subequations}
\begin{equation}
\hat S=\{(\hat N_F^2)_D^{(W)}\}^{dir}-\{(\hat N_F^2)_D^{(W)}\}^{ex},
\end{equation}
where
\begin{equation}
\{(\hat N_F^2)_D^{(W)}\}^{dir}=\{(\hat N_F)_D^{(W)}\}^2; \quad
\hat N_F=\sum_\alpha a_\alpha^\dagger a_\alpha,
\end{equation}
\begin{equation}
\{(\hat N_F^2)_D^{(W)}\}^{ex}=(\hat N_F)_D^{(W)}
-\sum_{\alpha\beta}(a_\alpha^\dagger a_\beta^\dagger)_D^{(W)}(a_\alpha a_\beta)_D^{(W)},
\end{equation}
\begin{equation}
\begin{array}{lll}
(a_\alpha^\dagger a_\beta^\dagger)_D^{(W)}&=&b_{\alpha\beta}^\dagger -\displaystyle\sum_{\gamma\delta}b_{\alpha\gamma}^\dagger b_{\beta\gamma}^\dagger b_{\gamma\delta},
\\
(a_\beta a_\alpha)_D^{(W)}&=&b_{\alpha\beta},
\\
(a_\alpha^\dagger a_\beta)_D^{(W)}&=&\displaystyle\sum_\gamma b_{\alpha\gamma}^\dagger b_{\beta\gamma}.
\end{array}\end{equation}
\end{subequations}
On the other hand, by mapping
\begin{equation}
(\hat N_F^2)^{dir}=\hat N_F^2,
\quad (\hat N_F^2)^{ex}=\hat N_F-\displaystyle\sum_{\alpha\beta}a_\alpha^\dagger a_\beta^\dagger a_\alpha a_\beta,
\end{equation}
with $U_\xi^{(W)}$, we obtain
\begin{subequations}
\begin{equation}
((\hat N_F^2)^{dir})_\xi^{(W)}=\hat T_B^{(W)}(\hat N_F)_D^{(W)}\hat T_B^{(W)}(\hat N_F)_D^{(W)}\hat T_B^{(W)},
\end{equation}
\begin{equation}
((\hat N_F^2)^{ex})_\xi^{(W)}=T_B^{(W)}(\hat N_F)_D^{(W)}\hat T_B^{(W)}-\hat T_B^{(W)}\displaystyle\sum_{\alpha\beta}(a_\alpha^\dagger a_\beta^\dagger)_D^{(W)} \hat T_B^{(W)} (a_\alpha a_\beta)_D^{(W)}\hat T_B,
\end{equation}
\end{subequations}
where we use the following: $(O_FO'_F)_\xi=(O_F)_\xi (O'_F)_\xi$ from ${U_{-\xi}^{(W)}}^\dagger U_\xi^{(W)}=\hat 1_F$, $\hat Z^{(W)}=(2\hat N_B^{(W)}-1)!!\hat T_B^{(W)}$ from $\hat Z^{(W)}=\lim_{N_{max}\rightarrow\infty}\hat Z^{(A)}$ and Eq. (\ref{eq:normopA}), and the equation obtained by replacing Eq. (\ref{eq:ofxi1}) with MWFS.
Since
\begin{subequations}
\label{eq:dtbw}
\begin{equation}
\label{eq:dtbw1}
[\hat T_B^{(W)}, (a_\alpha^\dagger a_\beta^\dagger )_D^{(W)}]=0,
\end{equation}
\begin{equation}
\hat T_B^{(W)}(a_\beta a_\alpha)_D^{(W)}\hat T_B^{(W)}=(a_\beta a_\alpha)_D^{(W)}\hat T_B^{(W)},
\end{equation}
\begin{equation}
[\hat T_B^{(W)}, (a_\alpha^\dagger a_\beta)_D^{(W)}]=0,
\end{equation}
\end{subequations}
hold \cite{JDF71}, we obtain
\begin{subequations}
\label{eq:nf2w}
\begin{equation}
((\hat N_F^2)^{dir})_\xi^{(W)}=\{(\hat N_F)_D^{(W)}\}^2\hat T_B^{(W)},
\end{equation}
\begin{equation}
\label{eq:nf2w2}
((\hat N_F^2)^{ex})_\xi^{(W)}=\{(\hat N_F)_D^{(W)}
-\sum_{\alpha\beta}(a_\alpha^\dagger a_\beta^\dagger)_D^{(W)}(a_\alpha a_\beta)_D^{(W)}\}\hat T_B^{(W)}.
\end{equation}
\end{subequations}
$(\hat N_F^2)=(\hat N_F^2)^{dir}=(\hat N_F^2)^{ex}$, $((\hat N_F^2)^{dir})_\xi^{(W)}-((\hat N_F^2)^{ex})_\xi^{(W)}=0$, and we obtain
\begin{equation}
\label{eq:stbw}
\hat S\hat T_B^{(W)}=0,
\end{equation}
which indicates that the physical state vectors in MWFS are eigenvectors of $\hat S$, and their eigenvalues are 0.
Since $\hat T_B^{(A)}=\hat T_B^{(W)}\breve 1_B^{(A)}$, by applying $\breve 1_B^{(A)}$ to the right side of Eq. (\ref{eq:stbw}), we obtain
\begin{equation}
\label{eq:stba}
\hat S\hat T_B^{(A)}=0.
\end{equation}
The equations replacing $W$ in Eqs. (\ref{eq:dtbw}) and Eqs. (\ref{eq:nf2w}) with A, and $((\hat N_F^2)^{dir})_\xi^{(A)}-((\hat N_F^2)^{ex})_\xi^{(A)}=0$ also hold.
On the other hand,
\begin{equation}
\hat S\hat T_B\neq 0\quad  (\hat W\neq 0).
\end{equation}
The reason is as follows: if $\hat W\neq 0$, then the commutation relations of the phonon creation and annihilation operators do not close in $\{t\}$, hence $O_FO'_F\neq O_F\hat T_F O'_F$. Also, $\hat T_B^{(A)}\hat T_B\neq \hat T_B$, $\hat T_B\hat T_B^{(A)}\neq \hat T_B$, therefore,
\begin{equation}
[\hat T_B, (a_\alpha^\dagger a_\beta^\dagger )_D^{(A)}]\neq 0,
\hat T_B (a_\beta a_\alpha)_D^{(A)}\hat T_B\neq (a_\beta a_\alpha)_D^{(A)}\hat T_B,
[\hat T_B, (a_\alpha^\dagger a_\beta)_D^{(A)}]\neq 0.
\end{equation}

If $\hat W=0$,
\begin{equation}
(\hat Z^{(A)})^\xi=(\hat Z+\hat Z')^\xi=(\hat Z)^\eta+(\hat Z')^\xi
\end{equation}
holds, and we obtain
\begin{subequations}
\label{eq:uud}
\begin{equation}
U_\xi^{(A)}=U_\xi + U'_\xi,
\end{equation}
\begin{equation}
U'_\xi =(\hat Z')^{\xi-\frac 12}\widetilde U',
\end{equation}
\begin{equation}
\widetilde U'=\sum_{N=0}^{N_{max}}\sum_{\bar t}\vert N; \bar t)\langle N;\bar t\vert,
\end{equation}
\end{subequations}
where the part containing $\{t\}$ and the part containing $\{\bar t\}$ are completely separated.
$T_B^{(A)}$ is also separated between $\{t\}$ and $\{\bar t\}$,
\begin{equation}
\hat T_B^{(A)}={U_{-\xi}^{(A)}}^\dagger U_\xi^{(A)}=\hat T_B+U'_{-\xi}{}^\dagger U'_\xi,
\end{equation}
which makes $\hat T_B^{(A)}\hat T_B=\hat T_B$ hold.
By applying $\hat T_B$ to the right of Eq. (\ref{eq:stba}), we obtain
\begin{equation}
\hat S\hat T_B=0\quad (\hat W=0).
\end{equation}
This is a natural consequence of the conclusion that when NAMD holds, MAT and MBT are identical for not just non-Hermitian but any type of boson mapping.
Furthermore, if the sort of phonon excitation modes and their number are appropriately restricted, the norm operator has no zero eigenvalues, and $\hat T_B=\breve 1_B$ holds, and we obtain
\begin{equation}
\hat S\breve 1_B=0\quad (\hat W=0),
\end{equation}
which indicates that $\hat S$ judges the ideal boson state vectors $\vert N; t)$ as also physical.

{The Park operator is effective for both MAT and MBT if and only if NAMD holds.
The conventional claim for the Park operator misunderstands NAMD and has deduced the incorrect conclusion without properly applying the results derived from NAMD.

The mechanism by which the Park operator determines that ideal boson state vectors are physical if and only if NAMD holds can also be derived by the following procedure:}
Using Eq. (\ref{eq:babbmu}), $\hat S$ is also expressed as
\begin{equation}
\hat S=4\sum_{\mu_1\mu_2}b_{\mu_1}^\dagger b_{\mu_2}^\dagger b_{\mu_1}b_{\mu_2}+\sum_{\mu_1\mu_2\mu'_1\mu'_2}Y(\mu'_2\mu_1\mu_2\mu'_1)b_{\mu_1}^\dagger b_{\mu_2}^\dagger b_{\mu'_1}b_{\mu'_2},
\end{equation}
which indicates that the ideal boson state vectors $\vert N; t)$ are {not} judged generally as physical. 
When the phonon excitation mode $\{t\}$, however, is appropriately chosen and the upper limit of the number of phonon excitations $N_{max}$ is properly, that $\hat S$ judges $\vert N; t)\quad (N\le N_{max})$ as physical by applying the conditions satisfied under NAMD, which has been, for the first time, derived by the norm operator method.
$Y(\mu'_2\mu_1\mu_2\mu'_1)$ satisfies the following relation,
\begin{equation}
\langle\langle\mu'_1\mu'_2\vert\mu_1\mu_2\rangle\rangle
=((\mu'_1\mu'_2\vert\mu_1\mu_2))-Y(\mu'_2\mu_1\mu_2\mu'_1),
\end{equation}
From Eq. (\ref{eq:normopA}), we obtain
\begin{equation}
\langle\langle\mu'_1\mu'_2\vert\mu_1\mu_2\rangle\rangle
=((\mu'_1\mu'_2\vert\hat Z^{(A)}\vert\mu_1\mu_2))
=3((\mu'_1\mu'_2\vert\hat T_B^{(A)}\vert\mu_1\mu_2))\quad (N_{max}\ge 2).
\end{equation}
Using these, we can express $\hat S^{(A)}=\breve 1_B^{(A)}\hat S\breve 1_B^{(A)}$ as
\begin{equation}
\label{eq:sa}
\hat S^{(A)}=\sum_{\mu_1\mu_2}\sum_{\mu'_1\mu'_2}
3((\mu'_1\mu'_2\vert (\breve 1_B^{(A)}-\hat T_B^{(A)})\vert\mu_1\mu_2))
b_{\mu_1}^\dagger b_{\mu_2}^\dagger b_{\mu'_1}b_{\mu'_2}.
\end{equation}
Since $\hat S=\lim_{N_{max}\rightarrow\infty}\hat S^{(A)}$, we obtain
\begin{equation}
\hat S=\sum_{\mu_1\mu_2}\sum_{\mu'_1\mu'_2}
3((\mu'_1\mu'_2\vert (\hat 1_B-\hat T_B^{(W)})\vert\mu_1\mu_2))
b_{\mu_1}^\dagger b_{\mu_2}^\dagger b_{\mu'_1}b_{\mu'_2},
\end{equation}
This means that, regardless of whether or not we restrict the number of phonon excitations, as long as we adopt all phonon excitation modes as boson excitations, the ideal boson state vectors $\vert N; \mu)$ are not physical.
When NAMD holds, $\hat W=0$, $\hat Z^{(A)}=\hat Z+\hat Z'$ and Eq. (\ref{eq:znnons2}) hold, so Eq. (\ref{eq:sa}) is expressed as
\begin{equation}
\label{eq:sanamd}
\hat S^{(A)}=\sum_{t_1t_2}\sum_{t'_1t'_2}
3((t'_1t'_2\vert (\breve 1_B-\hat T_B)\vert t_1t_2))
b_{t_1}^\dagger b_{t_2}^\dagger b_{t'_1}b_{t'_2}
+\sum_{\bar t_1\bar t_2}\sum_{\bar t'_1\bar t'_2}
((\bar t'_1\bar t'_2\vert (3\breve 1_B-\hat Z')\vert \bar t_1\bar t_2))
b_{\bar t_1}^\dagger b_{\bar t_2}^\dagger b_{\bar t'_1}b_{\bar t'_2},
\end{equation}
which indicates that the ideal boson state vectors $\vert N;\mu)$ are not physical just because NAMD holds.
However, if we choose the type of phonon excitation mode $\{t\}$ and the upper limit $N_{max}$ of the excitation number appropriately so that $\hat T_B=\breve 1_B$, then we obtain
\begin{equation}
\label{eq:sanamdprop}
\hat S^{(A)}=\sum_{\bar t_1\bar t_2}\sum_{\bar t'_1\bar t'_2}
((\bar t'_1\bar t'_2\vert (3\breve 1_B-\hat Z')\vert \bar t_1\bar t_2))
b_{\bar t_1}^\dagger b_{\bar t_2}^\dagger b_{\bar t'_1}b_{\bar t'_2}.
\end{equation}
Since
\begin{equation}
(\hat S-\hat S^{(A)})\vert N; \mu)=0 \quad (N\le N_{max}),
\end{equation}
holds, the ideal boson state vectors $\vert N; t)$ with these appropriate restrictions satisfy,
\begin{equation}
\hat S\vert N; t)=\hat S^{(A)}\vert N; t)=0,
\end{equation}
which indicates that the Park operator also judges that the ideal boson state vectors $\vert N;t)\quad (N\le N_{max})$ are physical.

\section{Summary}
\label{sum}
By using the norm operator method, which extends and corrects the conventional boson expansion theories, we compared the two mappings of the boson expansion theory: the so-called mapping after truncation (MAT) and the mapping before truncation (MBT).

MAT introduces the phonons and maps the fermion subspace spanned by the multi-phonon state vectors with properly selected excitation modes and excitation number one-to-one into a boson subspace consisting only of the bosons corresponding to the phonons adopted to the multi-phonon state vectors. On the other hand, MBT is a method that restricts the boson subspace obtained by mapping the whole fermionic space (MWFS) to a space with specific boson excitation modes.

We have made clear the following: MWFS takes the phonon excitation modes that are not adopted as the boson excitations in MAT as the boson excitations, which makes MBT generally different from MAT{, and MBT becomes an improper mapping that has no one-to-one correspondence with the fermion subspace}. When the phonon commutation relations are closed within the excitation modes adopted as the boson excitation in MAT, because of the elimination of the above difference, MBT coincides with MAT for all types: the Hermitian and the non-Hermitian.

We have also made it clear that {the conventional claim for the Park operator misunderstands NAMD and has deduced the incorrect conclusion without properly applying the results derived from NAMD.
}The Park operator gives a correct judge {both in MBT and in MAT} if and only if the commutation relations of the phonons are closed within the excitation mode adopted as boson excitations in MTB.

\let\doi\relax

\end{document}